\documentclass[12pt]{article}

\begin{document}

\title{\bf Comments on Two-Loop Four-Particle Amplitude in
Superstring Theory}

\author{Jun-Bao Wu \\School of Physics, Peking University \\
Beijing 100871, P. R. China\\ \\
Chuan-Jie Zhu\thanks{Supported in
part by fund from the National Natural Science Foundation of China
with grant Number
90103004.} \\
Institute of Theoretical Physics,
Chinese Academy of Sciences\\
P. O. Box 2735,  Beijing 100080, P. R. China}

\maketitle


\begin{abstract}
It is shown that the four-particle amplitude of superstring theory
at two loops obtained in \cite{AllZhu1,AllZhu3} is equivalent to
the previously obtained results in \cite{IengoZhu2, IengoZhu3,
Zhu}. Here the ${\bf Z}_2$ symmetry in hyperelliptic Riemann
surface plays an important role in the proof.
\end{abstract}


In some previous papers \cite{AllZhu1, AllZhu3, AllZhu2} we have
computed explicitly the two loop $n$-particle amplitude in
superstring theory for all $n\le 4$ by using the newly obtained
measure of D'Hoker and Phong \cite{DHokerPhong2, DHokerPhong3,
DHokerPhong4, DHokerPhong5} (for a recent review see
\cite{DHokerPhong1, DHokerPhong6}). The new measure of D'Hoker and
Phong is unambiguous and slice-independent. For all $n\le 3$ we
proved that the $n$-particle amplitudes are identically zero. A
simple expression was also obtained for the 4-particle amplitude
which is independent of the insertion points of the supermoduli.
Our explicit results beautifully verifies the result of D'Hoker
and Phong, especially for the non-zero four-particle amplitude.

A natural question is the relation with the previously obtained
result in \cite{IengoZhu2, IengoZhu3, Zhu} which had been proved
to satisfy all the conditions for a qualified 4-particle
amplitude. In particular, the explicit result has also been used
by Iengo \cite{Iengo} to prove the vanishing of the 2-loop
correction to the $R^4$ term \cite{GrossWitten}, in agreement with
the indirect argument of Green and Gutperle \cite{GreenGutperle},
Green, Gutperle and Vanhove \cite{Green2}, and Green and Sethi
\cite{GreenSethi} that the $R^4$ term does not receive
perturbative contributions beyond one loop. Recently, Stieberger
and Taylor \cite{Stieberger}  also used the result of
\cite{IengoZhu2, IengoZhu3, Zhu} to prove the vanishing of the
heterotic two-loop $F^4$ term. For some closely related works we
refer the reader to the reviews \cite{Green3, Kiritsis}.

In this paper we will show that the explicit results obtained by
using different chiral measure are actually equivalent. This
should be the case because they are all derived from first
principle. In the proof we used the ${\bf Z}_2$ symmetry of the
conformal field theory on hyperelliptic Riemann surface.

Our starting point is the chiral integrand obtained in
\cite{AllZhu3}:
\begin{equation}
{\cal A} = \langle : \partial X(q_1) \cdot (
\partial X(q_1) +
\partial X(q_2) ) : \prod_{i=1}^4 \hbox{e}^{i k_i \cdot X(z_i, \bar
z_i)} \rangle  \prod_{i=1}^4 { q -z_i\over y(z_i) } .
\end{equation}
As we said in \cite{AllZhu3}, we need to symmetrize in $q_{1,2}$
to obtain an explicitly $q$ independent result. This
symmetrization is justified as follows. First we have
\begin{eqnarray}
: \partial X(q_1) \cdot (
\partial X(q_1) +
\partial X(q_2) ) : & = &   {1\over 2}: ( (\partial X(q_1))^2
- (\partial X(q_2) )^2) :\nonumber \\
& & \hskip -4cm + {1 \over 2} :( \partial X(q_1) +
\partial X(q_2) ) \cdot ( \partial X(q_1) + \partial X(q_2) ) :
\nonumber \\
& = &  -(T(q_1) - T(q_2) )  \nonumber \\
& & \hskip -4cm + {1 \over 2} :( \partial X(q_1) +
\partial X(q_2) ) \cdot ( \partial X(q_1) + \partial X(q_2) ) : \,
,
\end{eqnarray}
where $T(q_i)= - { 1\over 2} :(\partial X(q_i))^2:$ is the stress
energy tensor. In the above, the last  term gives an $q$
independent result (see \cite{AllZhu3}) and the first  term will
give an vanishing contribution to the amplitude after integration
over $z_i$ as we can prove as follows.

By the ${\bf Z}_2$ symmetry of the conformal field theory on
hyperelliptic Riemann surface\footnote{See \cite{Dixon,
Zamolodchikov, Radul, Knizhnik2} for works on conformal field
theory on ${\bf Z}_n$ Riemann surface.} we have
\begin{equation}
\langle T(q_1) \prod_{i=1}^4 \hbox{e}^{i k_i \cdot X(z_i, \bar
z_i)} \rangle = \langle T(q_2) \prod_{i=1}^4 \hbox{e}^{i k_i \cdot
X(\tilde z_i, \tilde{\bar z}_i)} \rangle,
\end{equation}
where $\tilde z_i$ is the $Z_2$ transformed point of $z_i$, i.e.,
$\tilde z_i$ and $z_i$ are the two identical points on the two
different sheets of the hyperelliptic Riemann surface. By using
this result,  we have
\begin{eqnarray}
& & \int\prod_{i=1}^4 { {\rm d}^2 z_i } \, \prod_{i=1}^4{ q - z_i
\over y(z_i)} \,\langle T(q_1) \prod_{i=1}^4 \hbox{e}^{i k_i \cdot
X(z_i, \bar z_i)} \rangle \times (\hbox{Right Part}) \nonumber \\
& & \qquad = \int\prod_{i=1}^4 { {\rm d}^2 z_i } \, \prod_{i=1}^4{
q - z_i \over y(z_i)}\langle T(q_2) \prod_{i=1}^4 \hbox{e}^{i k_i
\cdot X(\tilde z_i, \tilde{\bar z}_i)} \rangle \times (\hbox{Right
Part}) \nonumber \\
& & \qquad = \int\prod_{i=1}^4 { {\rm d}^2 \tilde z_i } \,
\prod_{i=1}^4{ q - \tilde z_i \over y(\tilde z_i)}\langle T(q_2)
\prod_{i=1}^4 \hbox{e}^{i k_i \cdot X(\tilde z_i, \tilde{\bar
z}_i)} \rangle \times (\hbox{Right Part})
 \nonumber \\
& & \qquad = \int\prod_{i=1}^4 { {\rm d}^2 z_i } \, \prod_{i=1}^4{
q - z_i \over y(z_i)} \,\langle T(q_2) \prod_{i=1}^4 \hbox{e}^{i
k_i \cdot X(z_i, \bar z_i)} \rangle \times (\hbox{Right Part}) .
\end{eqnarray}
In the above we have used the ${\bf Z}_2$ invariance of the
integration measure ${\rm d}^2 z_i$ and in the last step we have
changed the dummy integration variables $\tilde z_i$'s back into
$z_i$'s. We note that in the above proof we have symbolically
denoted the contribution of the right part as $(\hbox{Right
Part})$. For specific amplitudes the above reasoning can be
justified completely. So effectively we can set $T(q_1) = T(q_2)$
in the chiral integrand, as we have used in \cite{AllZhu3}. The
chiral integrand is then given as follows:
\begin{eqnarray}
{\cal A}  & = &  {1\over 2} \langle : (\partial  X(q_1)+
\partial X(q_2) )^2 : \prod_{i=1}^4 \hbox{e}^{i k_i \cdot X(z_i, \bar
z_i)} \rangle  \prod_{i=1}^4 { q -z_i\over y(z_i) } \nonumber \\
& = & - {1\over 2} \langle (\partial X^\mu(q_1) + \partial
X^\mu(q_2)) k_i\cdot X(z_i) \rangle \nonumber \\
& & \times \langle (\partial X_\mu(q_1) +
\partial X_\mu(q_2)) k_j\cdot X(z_j) \rangle
\langle \prod_{i=1}^4 \hbox{e}^{i k_i \cdot X(z_i, \bar z_i)}
\rangle \prod_{i=1}^4 { q -z_i\over y(z_i) } \nonumber \\
& =  & - {1\over 2}\sum_{i,j=1}^4 k_i \cdot k_j {1\over
q-z_i}{1\over q-z_j} \, \langle \prod_{i=1}^4 \hbox{e}^{i k_i
\cdot X(z_i, \bar z_i)} \rangle \prod_{i=1}^4 { q -z_i\over y(z_i)
} \nonumber \\
& = &    { s (z_1z_2 + z_3 z_4) + t (z_1z_4 + z_2 z_3 ) + u (z_1
z_3 + z_2 z_4) \over 2 \prod_{i=1}^4  y(z_i)} \langle
\prod_{i=1}^4 \hbox{e}^{i k_i \cdot X(z_i, \bar z_i)} \rangle .
\end{eqnarray}

Now we recall the result obtained in \cite{IengoZhu2, IengoZhu3,
Zhu} for the 4-particle chiral amplitude. We have
\begin{eqnarray}
\tilde{\cal A} & = & \left\{ \langle  \partial X(q_1) \cdot
\partial X(q_2)  \prod_{i=1}^4 \hbox{e}^{i k_i \cdot X(z_i, \bar
z_i)}
\rangle \right. \nonumber \\
& & \left. -4 I_{\rm gh}(q) \langle \prod_{i=1}^4 \hbox{e}^{i k_i
\cdot X(z_i, \bar z_i)} \rangle \right\} \, \prod_{i=1}^4 { q -
z_i \over y(z_i)},
\end{eqnarray}
where
\begin{eqnarray}
I_{\rm gh}(q) & = & - {1\over 8}\left( \sum_{i=1}^6 {1\over q -
a_i} - 2 \sum_{i=1}^3{1 \over q - b_i} \right) \left(\sum_{i=1}^6{
1\over q - a_i} - \sum_{k=1}^4  {1\over q- z_k}\right) \nonumber
\\
&  & \hskip -1cm - { 1\over 32}\left( \sum_{i=1}^6 {1\over
(q-a_i)^2} - 2 \sum_{i<j}^6{1\over q-a_i} \, {1\over q -a_j} + 8
\sum_{i<j}^3 {1\over q-b_i} \, {1\over q-b_j}\right) ,
\end{eqnarray}
is the contribution from the ghost supercurrent. In the above
expression the $b_i$'s are the three ghost insertion points and
should be set to $a_i$ if we choose $a_{i=1,2,3}$ as the three
moduli to be integrated.

By using the result of \cite{AllZhu3}, we found that the
contribution to the chiral integrand from ${\cal X}_1 + {\cal
X}_6$ is given as follows:
\begin{eqnarray} {\cal A}_1 + {\cal A}_6 & = &   \prod_{i=1}^4 { q -z_i\over y(z_i) }
\left\{ \langle
 \partial X(q_1)  \cdot\partial X(q_2)
\prod_{i=1}^4 \hbox{e}^{i k_i \cdot X(z_i, \bar z_i)} \rangle
\right.
\nonumber \\
& & - \Big[  \partial_{q_1}G_2(q_1,q_2) +
\partial_{q_2}G_2(q_2,q_1)
 \nonumber \\
& &  \left.  +   ( G_2(q_1,q_2) + G_2(q_2,q_1) ) \sum_{k=1}^4 { 1
\over q - z_k} \Big] \langle \prod_{i=1}^4 \hbox{e}^{i k_i \cdot
X(z_i, \bar z_i)} \rangle. \right\}
\end{eqnarray}
The expressions that we will need are given as follows:
\begin{eqnarray}
& & G_2(q_1,q_2) + G_2(q_2,q_1)  =  - {1\over 2} \, \Delta_1(q) +
\left[ { 1\over q-p_1} \, { (q-p_2)(q-p_3) \over (p_1-p_2)
(p_1-p_3)} + ... \right] \nonumber \\
& & \qquad  = - {1\over 2} \, \Delta_1(q) +
\sum_{a=1}^3 { 1\over q - p_a}, \\
& & \partial_{q_1} G_2(q_1,q_2) + \partial_{q_2} G_2(q_2,q_1)  =
{3\over 8}\, \Delta_1^2(q) + { 1\over 4}\, \Delta_2(q) \nonumber
\\
& & \qquad + \left[ {1\over q-p_1}\, \left( {1\over q-p_2} +
{1\over q-p_3} - \Delta_1(q) \right)  { (q-p_2)(q-p_3) \over
(p_1-p_2) (p_1-p_3)} + ... \right] \nonumber \\
& & = {3\over 8}\, \Delta_1^2(q) + { 1\over 4}\, \Delta_2(q) -
\Delta_1(q) \sum_{a=1}^3 { 1\over q - p_a} + \sum_{a<b}^3 {1\over
(q-p_a)(q-p_b) } .
\end{eqnarray}
(See \cite{AllZhu1, AllZhu3} for details and notations.)  By
direct comparison we found the following equality:
\begin{eqnarray}
{\cal A}_1 + {\cal A}_6  & = & \left\{ \langle  \partial X(q_1)
\cdot
\partial X(q_2)  \prod_{i=1}^4 \hbox{e}^{i k_i \cdot X(z_i, \bar
z_i)}
\rangle \right. \nonumber \\
& & \left. + 4 I_{\rm gh}(q) \langle \prod_{i=1}^4 \hbox{e}^{i k_i
\cdot X(z_i, \bar z_i)} \rangle \right\} \, \prod_{i=1}^4 { q -
z_i \over y(z_i)},
\end{eqnarray}
if we identify $p_i = b_i$, the three ghost insertion points. We
note the important "$+$" sign in front of $4I_{\rm gh}$. This
shows clearly that ${\cal A}_1 + {\cal A}_6$ is not the complete
expression for the chiral integrand.

To prove that ${\cal A}$ and $\tilde{\cal A}$ give identical
scattering amplitude, we first note the following:
\begin{eqnarray}
   & & \langle \partial X(q_1) \cdot \partial X(q_2)
\prod_{i=1}^4 \hbox{e}^{i k_i \cdot X(z_i, \bar z_i)} \rangle
\nonumber \\
&  &  \qquad =  \langle \partial X(q_1) \cdot \partial X(q_2)
\rangle \langle  \prod_{i=1}^4 \hbox{e}^{i k_i \cdot X(z_i, \bar
z_i)} \rangle  \nonumber \\
& & \qquad - \sum_{i,j=1}^4 \langle \partial X^\mu(q_1) k_i\cdot
X(z_i) \rangle \langle \partial X_\mu(q_2) k_j\cdot X(z_i) \rangle
\langle \prod_{i=1}^4 \hbox{e}^{i k_i
\cdot X(z_i, \bar z_i)} \rangle \nonumber \\
&   & \qquad = \left\{ \langle \partial X(q_1) \cdot \partial
X(q_2) \rangle-  {1\over 4}  \sum_{i,j=1}^4k_i\cdot k_j
  {1\over q -z_i}\, {1\over q-z_j} \right.
  \nonumber \\
  & & \hskip  2cm \left.   + {1\over 4}  \sum_{i,j=1}^4 k_i \cdot k_j
f(q_1,z_i)f(q_1,z_j)  \right\} \langle \prod_{i=1}^4 \hbox{e}^{i
k_i \cdot X(z_i, \bar z_i)} \rangle \, , \label{eqxx}
\end{eqnarray}
by using the following explicit expression for the $\langle
\partial X(q_{1,2}) X(z_i,\bar z_i)\rangle$ correlator:
\begin{equation}
\langle \partial X_\mu(q_a) X_\nu(z_i,\bar z_i)\rangle =
-{\eta_{\mu\nu}\over 2} \left( { 1\over q_a - z_i } +
f(q_a,z_i)\right) \, , \label{eqpxx}
\end{equation}
where
\begin{equation} f(q_a,z_i)   =  { 1\over   T} \int  {
y(z_i)\over y(q_a) } \, { 1\over q_a - z_i}    { (q_a-u)(q_a-v)
\over (z_i-u)(z_i-v) } \, \left| u-v \over y(u)y(v) \right|^2 {\rm
d}^2 u{\rm d}^2 v .
\end{equation}
From eq.~(\ref{eqxx}), we see that the second term will give half
of the result of ${\cal A}$ when we substitute it into
$\tilde{\cal A}$. It remains to compute the rest terms in
$\tilde{\cal A}$. We denote it as ${\cal A}_D$ and it is given as
follows:
\begin{eqnarray}
{\cal A}_D& = & \Big\{ \langle \partial X(q_1) \cdot \partial
X(q_2) \rangle +  {1\over 4} \sum_{i,j=1}^4 k_i \cdot k_j
f(q_1,z_i)f(q_1,z_j) \nonumber \\
& & \qquad - 4 I_{\rm gh}(q) \Big\} \langle \prod_{i=1}^4
\hbox{e}^{i k_i \cdot X(z_i, \bar z_i)} \rangle
 \, \prod_{i=1}^4 { q - z_i \over y(z_i)}\, .
\end{eqnarray}
As we proved in \cite{IengoZhu2, IengoZhu3, Zhu}, the above
expression is independent of $q$ after integration over all
moduli. To simplify the computation we can make a convenient
choice for the insertion points $q_{1,2}$ or $q$.  We will take
the limit $q_1 \to z_1$. In this limit we have
\begin{eqnarray}
{\cal A}_D   & = &   {z_{12}z_{13}z_{14} \over \prod_{i=1}^4
y(z_i)} \left\{ {1\over 2} \sum_{ j=2}^4 k_1 \cdot k_j
 f(z_1,z_j) \right. \nonumber \\
 & &  \qquad \left.
 + { 1\over 2} \left( \Delta_1(z_1) - 2 \sum_{i=1}^3 { 1
\over z_1 - b_i}\right)  \right\} \langle \prod_{i=1}^4
\hbox{e}^{i k_i \cdot X(z_i, \bar z_i)} \rangle   . \label{eqad}
\end{eqnarray}
We will prove that the second term vanishes and the first term
gives the other half of ${\cal A}$. We note that this is the case
only after we make this special choice for $q$.

First we start with the second term. By using the result of
\cite{IengoZhu3}, we can change the integration over the moduli
$a_{1,2,3}$ and $z_{1,\cdots,4}$ into $a_{1,\cdots,6}$ and $z_1$
by fixing $z_{2,3,4}$. The chiral integrand doesn't change and it
is still given by the following expression:
\begin{equation}
{\cal A}_{D2}     =    {z_{12}z_{13}z_{14} \over 2 \prod_{i=1}^4
y(z_i)}  \left( \sum_{i=1}^6 {1\over z_1 - a_i}   - 2 \sum_{i=1}^3
{ 1 \over z_1 - b_i}\right)  \langle \prod_{i=1}^4 \hbox{e}^{i k_i
\cdot X(z_i, \bar z_i)} \rangle .
\end{equation}
The important point is the following: the rest part of the
integrand is invariant under modular transformation, i.e.,
symmetric under $a_i \leftrightarrow a_j$. (We note that $b_i =
a_i$.) On the other hand, ${\cal A}_{D2}$ is antisymmetric under
the following modular transformation:
\begin{equation}
a_1 \to a_4 , \qquad a_2 \to a_5 , \qquad a_3 \to a_6.
\end{equation}
So after integration over all $a_i$'s, ${\cal A}_{D2}$ gives a
vanishing result. This result can also be understood from another
point of view. Because of modular invariance, each term in the
summation in ${\cal A}_{D2}$ gives the same result after
integration over all $a_i$'s and all of them add to zero.

The other term in (\ref{eqad}) is:
\begin{equation}
 {\cal A}_{D1}  =  {z_{12}z_{13}z_{14} \over 2\,
 \prod_{i=1}^4 y(z_i)}
  \sum_{ j=2}^4 k_1 \cdot k_j
 f(z_1,z_j)   \langle \prod_{i=1}^4
\hbox{e}^{i k_i \cdot X(z_i, \bar z_i)} \rangle  . \label{eqd1}
\end{equation}
In order to compute it explicitly, we first note the result of
$T(q_1) = T(q_2)$ proved before. Explicitly we have\footnote{All
equality, such as $T(q_1) = T(q_2)$,  should be understood as
expressions inserted in the complete amplitude and integration
over all moduli is implicit.}:
\begin{eqnarray}
0 & = & \langle (T(q_1) - T(q_2)) \prod_{i=1}^4 \hbox{e}^{i k_i
\cdot
X(z_i, \bar z_i)} \rangle \nonumber \\
& =  &    {1\over 4} \sum_{i,j=1}^4 k_i \cdot k_j \, \left\{ { 1
\over q-z_i} f(q_1,z_j) \right. \nonumber \\
& & \qquad  \left. +  { 1 \over q-z_j} f(q_1,z_i) \right\} \,
\langle \prod_{i=1}^4 \hbox{e}^{i k_i \cdot X(z_i, \bar z_i)}
\rangle \, ,
\end{eqnarray}
by using the explicit result for the $\langle \partial X(q_a)
X(z_i)\rangle$ correlator given in (\ref{eqpxx}). If we set $q_1
\to z_1$, we have
\begin{eqnarray}
& & \langle (T(q_1) - T(q_2))  \prod_{i=1}^4 \hbox{e}^{i k_i \cdot
X(z_i, \bar z_i)} \rangle \nonumber \\
& & \qquad \to {1\over 2} \left\{   {1\over q - z_1} \,
\sum_{j=2}^4 k_1 \cdot k_j \, f(z_1,z_j) \right. \nonumber \\
& & \qquad \qquad  \left. +
 {1\over q - z_1} \, \sum_{j=2}^4
  { k_1\cdot k_j\over z_1 - z_j} \right\}
 \, \langle \prod_{i=1}^4 \hbox{e}^{i k_i
\cdot X(z_i, \bar z_i)} \rangle =0  \, . \label{diff}
\end{eqnarray}
Substituting this result into eq.~(\ref{eqd1}), we have
\begin{eqnarray}
{\cal A}_{D1} & = & - {z_{12}z_{13}z_{14} \over 2\,
 \prod_{i=1}^4 y(z_i)}
  \sum_{ j=2}^4 {k_1 \cdot k_j \over z_1 - z_j}
 \langle \prod_{i=1}^4
 \hbox{e}^{i k_i \cdot X(z_i, \bar z_i)} \rangle \nonumber \\
 & = &  { s (z_1z_2 + z_3 z_4) + t (z_1z_4 + z_2 z_3 ) + u (z_1
z_3 + z_2 z_4) \over 4 \prod_{i=1}^4  y(z_i)} \nonumber \\
& & \times \langle \prod_{i=1}^4 \hbox{e}^{i k_i \cdot X(z_i, \bar
z_i)} \rangle = { {\cal A} \over  2 } .
\end{eqnarray}
This proves
\begin{equation}
{\cal A} = \tilde{\cal A}.
\end{equation}

\section*{Acknowledgments}

Chuan-Jie Zhu would like to thank Roberto Iengo for suggesting the
problem, reading the paper and comments. He would also like to
thank E. D'Hoker and D. Phong for discussions.

\end{document}